# The First Orbital Flight of the ELROI Optical Satellite License Plate

David M. Palmer, Rebecca M. Holmes, Charles T. Weaver
Los Alamos National Laboratory
PO Box 1663, Los Alamos, NM 87545, USA; 505 665-6863
palmer@lanl.gov

**ABSTRACT**

Space Object Identification is one of the cornerstones of Space Traffic Control and a requirement for successful operation of a spacecraft.

ELROI, the *Extremely Low Resource Optical Identifier*, is a new concept that can provide a self-powered satellite identification beacon in a package the size of a thick postage stamp. Its small size, low cost, and fully autonomous operation make it usable by all space objects, including CubeSats and inert debris objects.

The beacon's signal is received on the ground using a small telescope equipped with a photon-counting detector which can unambiguously determine the satellite identification number during a single pass overhead. Additional information can be included in the signal to aid in anomaly diagnosis and resolution, further improving spaceflight reliability and safety.

The first ELROI unit in orbit was launched December, 2018 as a payload on the student CubeSat NMTSat. We are now searching for the identification signal.

## INTRODUCTION

Space is big.[1] But the most useful parts of space are getting crowded with thousands of active and retired satellites, cast-off rocket bodies, fragments, and other pieces of debris. It is getting worse, with multiple constellations of thousands of satellites each being prepared for launch. Space Traffic Management (STM) tries to keep track of every object in space to avoid collisions and other conflicts, and an important component of this is Space Object Identification (SOI). SOI requires identifying the object shortly after launch and then making observations on a sub-weekly timescale to maintain custody.

Beyond the STM application, SOI is required for effective operations of satellites, especially in the perilous early days just after launch. A single rocket can now release more than a hundred CubeSats.[2] After launch, distinguishing which one is which can be a major challenge. Of the 65 objects orbited by the SSO-A launch on December 3, 2018,[3] 19 are still unidentified and listed as OBJECT *XX* in the space-track.org catalog more than four months later.

This can make early post-launch operations difficult, as ground stations do not know which object to point their antennas at for commanding and telemetry. Even in more traditional launches, a single payload and the rocket that carried it have been confused, leading to an inability to command a spacecraft to open its solar panels before its batteries ran out.[4]

Satellites can help to identify themselves. When a July 2017 Proton launch orbited 73 objects,[5] 55 satellites built for two constellations (47 from Planet, 8 from Spire) quickly determined their own orbits using GPS and reported them to the ground, simplifying identification. Of the remaining 18 tracked objects only 9 have been identified, including several others that also had on-board orbit determination and reporting.

An identification beacon definitely helps the SOI problem. If the beacon signal is incorporated into the primary radio transmitter, then the beacon can only be received if the frequency is known, which is not necessarily the case for an unidentified object. If a standard for dedicated radio beacons were developed, then a standardized beacon frequency could be used, but an additional dedicated radio transmitter beacon would tend to be bulky, heavy, power-hungry, and expensive (at least compared to a minimal CubeSat). To reduce radio frequency interference (RFI), radio transmitters are turned off at the end of a satellite's operational life. Thereafter, continual observation of a satellite is required to maintain track custody and avoid losing the satellite's identification.

ELROI, which stands for *Extremely Low Resource Optical Identifier*, is an optical beacon which can be developed into a package that consumes very little from its host spacecraft.[6,7,8,9] It can be made into a small, lightweight package (a few cubic centimeters, a few grams) that operates autonomously using its own solar power system. It places no power or data requirements on its host so it can be attached even to inert objects such as rocket bodies and interstages. Because it is an optical beacon it produces no RFI and can continue operating beyond the end of the mission. The simplicity of its circuitry enables low-cost production that will allow it to be attached or built in to everything that goes into space.

The first ELROI flight prototype was installed in NMTSat, a 3U CubeSat built by the New Mexico Institute of Mining and Technology. It was launched on 2018 December 16 on the ELaNa XIX flight from New Zealand on a Rocket Lab Electron. This paper describes the first attempts to receive the ELROI signal from this satellite.

**THE ELROI CONCEPT**

ELROI is described in much more detail in Palmer & Holmes[7]. Some of this summary is abbreviated or taken directly from that and other papers[6,8,9].

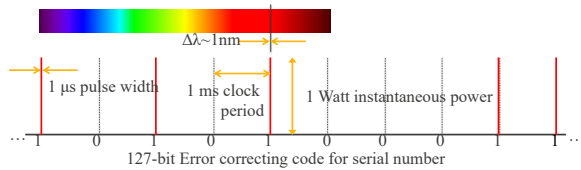

**Figure 1 The ELROI signal—brief, bright pulses of monochromatic light on a periodic clock interval with a repeating error-correcting code—allows extreme background rejection and permits a very low power signal to be recovered despite otherwise overwhelming background.**

The ELROI signal is produced by high-power laser diodes that are diffused to emit over a wide angular range so that they can be seen from anyplace visible on the ground. These diodes emit very short flashes of light with precise timing hundreds of times per second that encode a satellite's serial number. Because the flashes are so brief, the lasers are off 99.9% of the time and on average draw only a few milliwatts of power, which can be supplied by a few square centimeters of solar cell with a battery (the size you would find in wrist-watch) to keep operating through orbital night. The low power, and the simplicity of the required circuitry, will allow the beacon to be packaged as low cost, small, lightweight self-contained unit that can fit the size, weight, power, integration, and economic budget of even the smallest CubeSat.

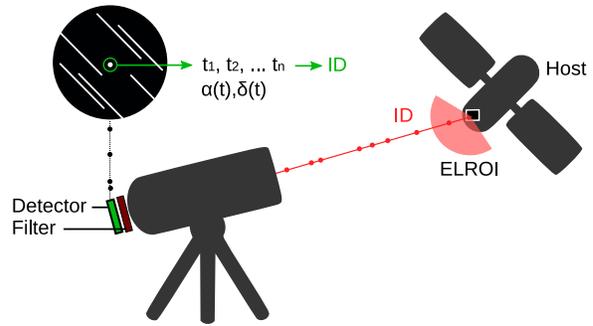

**Figure 2 Overview of the ELROI system. The beacon is attached to a satellite and continuously emits its optical signal—encoding a unique ID number—over a wide solid angle. A ground telescope collects a small portion of the emitted photons, which are detected with a photon-counting sensor. A narrowband filter centered on the beacon wavelength rejects background light. The recorded data (circular inset) consists of a list of photon detection times at a tracked location (green circle). Streaks represent background stars. The data analysis technique uses the timing characteristics of the ELROI signal to eliminate more than 99% of background photons, making it possible to read the ID in a single pass even if the signal is only a few photons per second.**

The signal is received by pointing a small telescope at the satellite and using a single-photon detector to recover the timing of ELROI's emitted light. Our ground station is a 35-cm aperture telescope (a size used by serious amateur astronomers) with a specialized detector developed by our group at Los Alamos National Laboratory (although commercial equivalents are available). This system is appropriately sized to determine the ELROI serial number during a single overhead pass of a LEO satellite with several minutes of observation. Engineering trade-offs can be made between size and pointing accuracy of the telescope, complexity and efficiency of the detector, orbital regime, observation time, and telescope location.

The timing, coding, and spectral characteristics of the signal allow extreme background rejection to distinguish the ELROI photons from sunlight bouncing off the satellite and determine the satellite identification number with very high reliability.

Successful ground tests[7] validated the concept at a range of 15 km using attenuators and reduced-size optics to scale the signal to that expected at satellite-to-ground distances. This demonstrated that ELROI signals of a few photons per second can be extracted from data taken at much higher background levels and that the encoded registration number can be confidently identified.

The ELROI signal is intentionally open and accessible to anyone with a ground station. The ID of each beacon will be stored in an open registry, along with contact information for its operator and other information. This allows ELROI to be adopted as an international standard, read by ground stations around the world to assist in the worldwide problem of STM. The beacon can transmit additional data beyond the ID, giving satellite operators a backup channel for anomaly resolution and other diagnostic purposes. This, along with the benefits to the spacecraft operator of being able to identify their own satellite, can drive adoption of ELROI even in the absence of international norms or mandates.

**ELROI ON NMTSAT**

Following the successful ground tests, we solicited flight opportunities. The first of these was offered by the New Mexico Institute of Mining and Technology (NMT) which was in the process of developing their first educational satellite, NMTSat.[10]

NMTSat is a 3U CubeSat in a Pumpkin CubeSat Kit bus, with electronics built as a stack of PC-104 form-factor boards. Attitude control is provided by a passively-damped magnet that tends to keep one of the short axes of the spacecraft aligned with the local magnetic field lines in orbit. Thus one of the 3Ux1U faces tends to point generally North and, when over the Northern temperate regions, somewhat downwards.

An ELROI unit (Figure 2) was built on a PC-104 card and incorporated into the NMTSat electronics stack. This contained two optical diffusers exposed through the North and South faces of the satellite. Each diffuser distributed light from two laser diodes in a 120° wide cone. The North diffuser has two red ($\lambda$=638 nm) laser diodes pulsed at 700 mW and 1000 mW instantaneous peak power; while the South diffuser has a red ($\lambda$=638 nm @ 1000 mW) and a blue ($\lambda$=450 nm @ 1600 mW) laser diode. Each diode is pulsed with its own identification code and with staggered timing to allow clean separation of the signals. The variety of diodes allows us to evaluate different diode manufacturers and wavelengths.

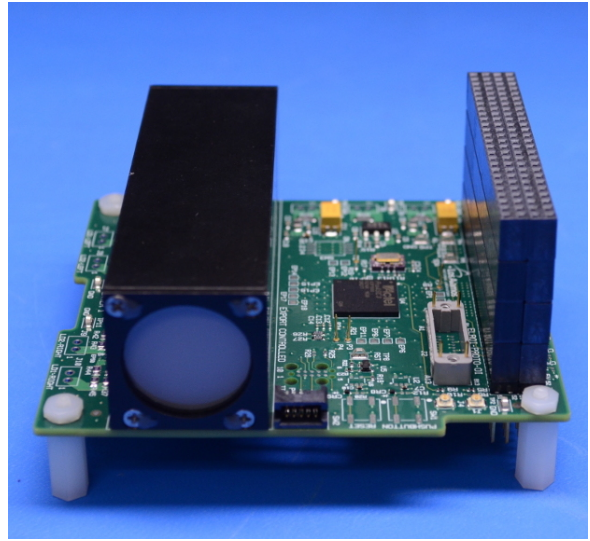

**Figure 3 The ELROI-PC104 unit that was installed on NMTSat.**

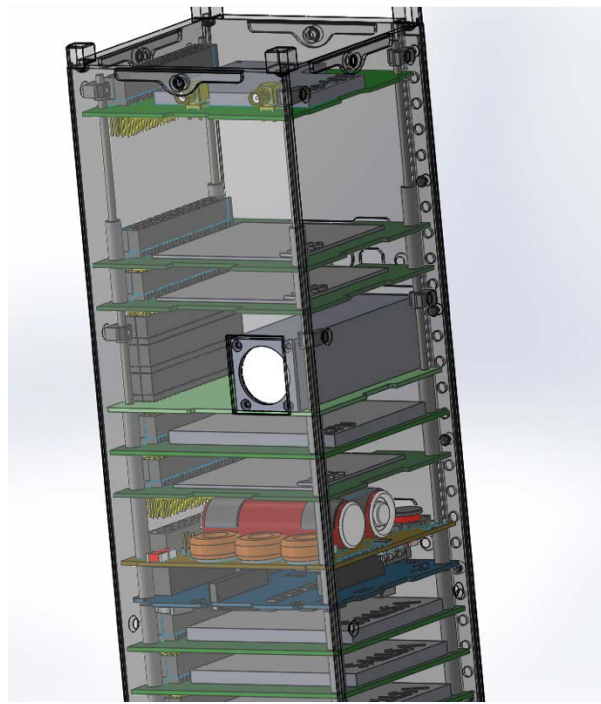

**Figure 4 Installation location of ELROI in NMTSat.**

The ELROI unit was connected to the spacecraft power system so that it would turn on as soon as the spacecraft was energized and start to emit its optical signal after a 45-minute delay (as required by the P-POD standard). Thus it can operate without a direct command from the spacecraft processor. However, the spacecraft processor can command it to change parameters (such as the clock interval, duty cycle, and identification code) away from the power-on default. During a nominal boot sequence, the NMTSat processor commands several parameter changes allowing a remotely accessible diagnostic of whether the processor completed booting or, if the ELROI signal is not in the expected end state, where in the boot sequence it failed.

## LAUNCH AND CONTACT AND OBSERVATION ATTEMPTS

NMTSat was launched as part of NASA's ELaNa XIX program on a Rocket Lab Electron launcher from New Zealand on 16 December 2018. The launcher carried 13 satellites including NMTSat. However, satellite tracking released by the US Air Force 18th Space Control Squadron (on www.space-track.org ) indicated 15 objects in orbit (catalog numbers #43849-#43863; international designator 2018-104) including the rocket's upper stage which was propulsively moved to a lower orbit after satellite deployment and has since re-entered.

Of the 14 remaining orbital objects, 7 have been contacted by radio and identified with specific satellites. NMTSat was not one of the contacted satellites and so it is one of the 6 remaining satellites that may be among the 7 unattributed tracks. Some of these tracks were tentatively identified with specific objects, but these identifications were later withdrawn. Our observation program has been spread across all of these tracks to try to identify the ELROI signal.

The 85° orbital inclination of these objects results in a slow precession of the orbital plane relative to the Sun-Earth line. As a result, nighttime passes of the objects over our ground station occurred while the satellites were in eclipse (not illuminated by the Sun and thus not detectable with our telescope) during the first three months after launch.

Observations of each object involved pointing the telescope using predictions based on orbital elements (TLEs) provided by space-track.org and digitizing all photons in the detector's 0.5° diameter Field Of View (FOV). While the satellite was in eclipse, the location of the satellite relative to the FOV was poorly-constrained due to accuracy limitations in the available orbital elements. To search the entire FOV, data was divided into time intervals and subregions within the FOV, and each block of data was analyzed to look for the ELROI signal. The time interval duration and region size were varied in these searches to trade sensitivity against tracking and clock offset behaviors.

In many cases, the TLEs made available soon after the observation are sufficiently different from the TLEs used to point the telescope, often to the extent that they predict that the satellite was not in our FOV during the observation.

Beginning in late March, 2019, orbital precession provided overhead passes where the satellite was illuminated by the Sun when our ground station was sufficiently dark to make observations (known as 'terminator conditions'). During these passes, each satellite can be seen directly when it is in our detector's FOV. If the satellite is not visible, then the telescope pointing can be adjusted to seek along the predicted track to find the satellite, and guidance corrections can be made throughout the pass to keep it in the FOV. During analysis only data from the satellite location need be analyzed, greatly improving the sensitivity and reliability of the search.

## OBSERVATION RESULTS

ELROI has not been detected in any of the data collected so far from the 7 unidentified tracks. The quality of the data available varies from track to track. Some of the observations can be used to narrow the remaining search. For example, the track named Object C (#43851) in the space-track.org catalog is much brighter (10-100×) than the others, suggesting ~1 m$^2$ of reflecting area, inconsistent with a 3U CubeSat. Objects E, H and K (#43853, #43856, #43858) show periodic brightness variations indicating that they are spinning with a period of a few seconds. This is typical for CubeSats that have no attitude control system, but is unlikely for NMTSat with its passive magnetically damped stabilization.

The non-detection of ELROI in the remaining objects is not yet sufficiently stringent to rule out a properly-functioning ELROI unit on one of them. This would require observations under clear dark skies, in terminator with the satellite in the FOV, while in the Southern sky as observed from our ground station (assuming the magnetic stabilization points the 120° cone of the ELROI diffuser to the North and downwards towards our ground station).

## POTENTIAL CAUSES OF NON-DETECTION

It is possible that the ELROI unit on NMTSat is fully functional and that none of the current observations of whichever track is NMTSat are sufficiently good to detect the signal.

The observations may be good enough, but the signal will only show up with better analysis. The data analyses that we have made can be refined in several ways to improve our sensitivity to ELROI. We can analyze data assuming an ensemble of orbital elements consistent with the expected error ranges of the TLE, using the location of stars in the telescope FOV to remove tracking error. Incorporating predicted satellite range would allow longer signal integrations and greater sensitivity by correcting for changes in the doppler shift of the signal clock. The analysis sensitivity can also be improved by taking into account the variation in expected signal strength vs viewing angle relative to the presumed alignment of satellite along the magnetic field lines.

The parameters of the ELROI signal may have changed due to the effects of the space environment, or due to errant reconfiguration commands from the spacecraft processor. If the temperature of the laser diode is outside of the expected range, this can cause its wavelength to shift outside of the bandpass provided by the filter on our ground station.

Finally, a malfunction either in the spacecraft systems or the ELROI unit itself may prevent it from functioning.

We will continue to make observations of the unidentified tracks, improve the analysis software, and attempt to make radio contact with NMTSat.

**PLANNED ELROI DEVELOPMENT**

Even if the NMTSat ELROI is never detected, we are advancing our development work.

We have delivered two further ELROI units for satellites scheduled for launch in late 2020. We have produced additional ELROI units and are actively looking for future launch opportunities. We are also recruiting additional ground stations to search for ELROI on NMTSat, and on future satellites. We continue to develop designs and concepts for ELROI.

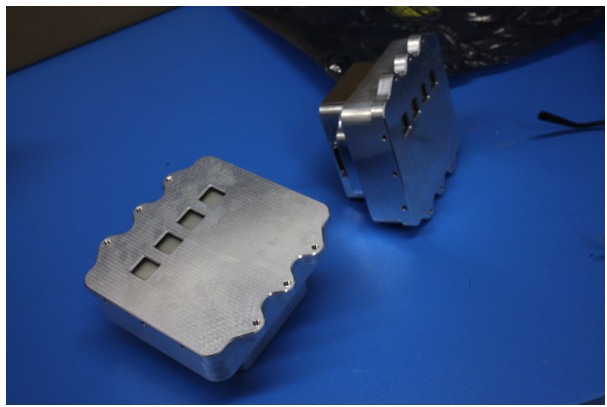

**Figure 5  Two ELROI units delivered for a launch in 2020.**

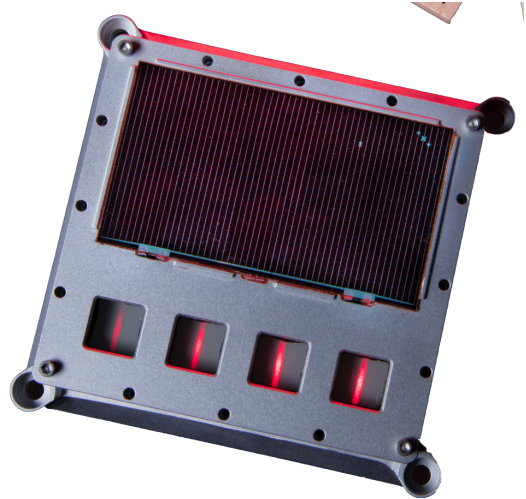

**Figure 6  Autonomous solar-powered ELROI units available for future flights.**

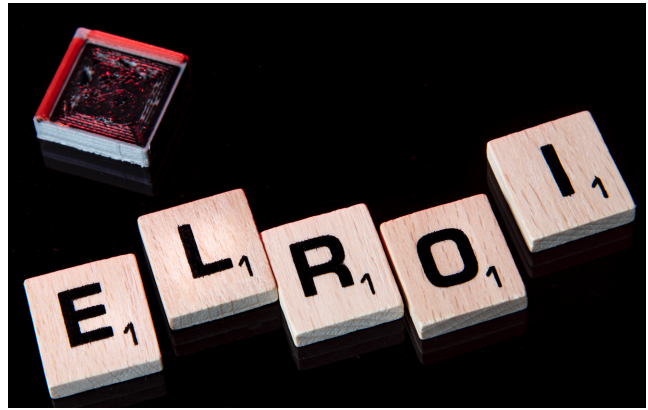

**Figure 7 An autonomous solar-powered ELROI unit could be manufactured with a size of 2x2x0.5 cm$^3$, about the size of a scrabble tile (also pictured).**

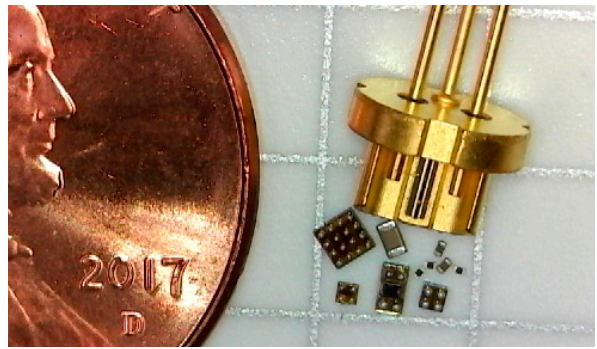

**Figure 8 The additional electronic components required for a built-in spacecraft-powered ELROI unit are shown here.
5 mm$^2$ grid and US 1¢ coin included for scale.**

Figures 5-8 show current and planned development stages for ELROI. The current flyable prototype units take up about 1/3 of a 1U CubeSat's volume and mass. Straightforward engineering can produce a fully autonomous unit (needing no power, commanding, or data support from its host) with a volume of a few cm$^3$ and a weight of a few grams. ELROI capability can also be designed-in to a spacecraft by adding a small number of very small components.

## CONCLUSIONS

It is too soon to know whether the first ELROI orbital test was successful. We continue to try to detect its signal.

ELROI units have been delivered for a late-2020 launch. We are also looking for further flight opportunities. Please contact the author at palmer@lanl.gov if you wish to fly an ELROI unit on your spacecraft.

We are also soliciting participations by other ground stations. Satellite Laser Ranging incorporates all the hardware and expertise required to read an ELROI beacon, and other optical satellite tracking applications can be augmented with the addition of a suitable detector system.

Potential manufacturers of ELROI units are also invited to contact us. The design complexity is well within the capabilities of even the smallest space electronics firm.

After ELROI has been demonstrated on orbit, we expect consultation with stakeholders to develop optimal choices for wavelength, clocking, and other signal characteristics that can be adopted as a world-wide standard.

The eventual goal is to put an ELROI beacon on everything that goes into space, solving the Space Object Identification problem in our increasingly crowded skies.

## ACKNOWLEDGEMENTS

Initial work on this project was supported by the U.S. Department of Energy through the Los Alamos National Laboratory (LANL) Laboratory Directed Research and Development program as part of the IMPACT (Integrated Modeling of Perturbations in Atmospheres for Conjunction Tracking) project. Further work was supported by the Richard P. Feynman Center for Innovation at LANL. ELROI hardware and software was developed at LANL by Louis Borges, Richard Dutch, Darren Harvey, David Hemsing, Joellen Lansford and Charles Weaver, with thermal analysis by Alexandra Hickey, Lee Holguin, and Zachary Kennison. The NMTSat team at the New Mexico Institute of Mining and Technology in Socorro, NM is Sawyer Gill, James Z. Harris, Joellen Lansford, Riley Myers, Aaron Zucherman, and Anders M. Jorgensen.

*Cleared for unlimited release LA-UR-19-25352.*